\documentclass[%
 reprint,
superscriptaddress,
 amsmath,amssymb,
]{revtex4-1}

\usepackage{bbold}
\usepackage{mathptmx}
\usepackage{subfig}
\usepackage{psfrag,graphicx}
\usepackage{dcolumn}
\usepackage{amsmath,amssymb}
\usepackage{bm}
\usepackage{color}
\usepackage{latexsym}
\usepackage{epstopdf}
\usepackage{color}
\usepackage[english]{babel}
\usepackage{latexsym}
\usepackage{psfrag,graphicx}
\usepackage{amsmath}
\usepackage{amssymb}
\usepackage{amsfonts}
\usepackage{bm}
\usepackage{natbib}
\usepackage{epstopdf}
\usepackage{appendix}
\usepackage{rotating}
\usepackage[english]{babel}
\usepackage{aeguill}
\usepackage{ulem}
\usepackage[justification=justified]{caption}

\definecolor{mygrey}{gray}{0.35}
\definecolor{myblue}{rgb}{0.2,0.2,0.8}
\definecolor{myzard}{cmyk}{0,0,0.05,0}
\definecolor{mywhite}{rgb}{1,1,1}
\definecolor{mywhite}{rgb}{1,1,1}
\definecolor{myred}{rgb}{1,0.,0.3}

\usepackage[colorlinks=true,citecolor=myblue,linkcolor=myblue]{hyperref}

\def\ba{\begin{align}}
\def\enda{\end{align}}
\def\bi{\begin{itemize}}
\def\ei{\end{itemize}}

\def\be{\begin{equation}}
\def\ee{\end{equation}}
\def\bea{\begin{eqnarray}}
\def\eea{\end{eqnarray}}
\def\bse{\begin{subequations}}
\def\ese{\end{subequations}}

\newcommand{\ket}[1]{|{#1}\rangle}                       % ket
\newcommand{\average}[1]{\langle {#1} \rangle}           % media < >

\newcommand{\Ignore}[1]{ }

\def\i{\text{i}}

\begin{document}

\preprint{APS/123-QED}

\title{Thermodynamic Limit in the Two-qubit Quantum Rabi Model with Spin-Spin Coupling}

\author{R. Grimaudo}
\address{Department of Physics and Astronomy ``E. Majorana", University of Catania, Via S. Sofia, 64 I-95123 Catania, Italy}

\author{G. Falci}
\address{Department of Physics and Astronomy ``E. Majorana", University of Catania, Via S. Sofia, 64 I-95123 Catania, Italy}
\address{CNR-IMM, UoS Universit\`a, 95123, Catania, Italy}
\address{INFN Sez. Catania, 95123 Catania, Italy}

\author{A. Messina}
%\address{INFN, Sezione di Catania, I-95123 Catania, Italy}
\address{ Department of Mathematics and Informatics, University of Palermo, Via Archirafi 34, I-90123 Palermo, Italy}

\author{E. Paladino}
\address{Department of Physics and Astronomy ``E. Majorana", University of Catania, Via S. Sofia, 64 I-95123 Catania, Italy}
\address{CNR-IMM, UoS Universit\`a, 95123, Catania, Italy}
\address{INFN Sez. Catania, 95123 Catania, Italy}

\author{A. Sergi}
\address{Dipartimento di Scienze Matematiche e Informatiche, Scienze Fisiche e Scienze della Terra, Universit\`{a} degli Studi di Messina, Viale F. Stagno d'Alcontres 31, 98166 Messina, Italy}
\address{Institute of Systems Science, Durban University of Technology, P.O. Box 1334, Durban 4000, South Africa}

\author{E. Solano}
\address{Kipu Quantum, Greifswalderstrasse 226, 10405 Berlin, Germany}

\author{D. Valenti}
\address{Department of Physics and Chemistry ``Emilio Segr\`{e}",
University of Palermo, viale delle Scienze, Ed. 18, I-90128, Palermo, Italy}

\date{\today}

\begin{abstract}
 
The occurrence of a second-order superradiant quantum phase transition is brought to light in a quantum system consisting of two interacting qubits coupled to the same quantized field mode. We introduce an appropriate thermodynamic-like limit for the integrable two-qubit quantum Rabi model with spin-spin interaction. Namely, it is determined by the infinite ratios of the spin-spin and the spin-mode couplings to the mode frequency, regardless of the spin-to-mode frequency ratios.

\end{abstract}



\maketitle

%\Ignore{
\textit{Introduction.}
Over the past decades quantum phase transitions \cite{Sondhi97,Laughlin98,Sachdev99} have attracted a great deal of attention in the condensed matter community \cite{Vojta03}.
%Experimental and theoretical developments have clearly shown that many unsolved puzzles in condensed matter systems, such as rare-earth magnetic insulators [5], heavy-fermion compounds [6, 7], high-temperature superconductors [8, 9], and two-dimensional electron gases [1, 10], can be explained and understood on the basis of the existence of zero-temperature quantum critical points \cite{Vojta03}.
Traditionally, quantum phase transitions (QPTs) are intended to occur in the thermodynamic limit, that is when the number of elements of the system is very large \cite{Sachdev99, Vojta03}.
However, critical phenomena, for instance jumps of some physical observable, can also occur in systems with few degrees of freedom \cite{Bakemeier12, Ashhab13}.
In this case, one speaks of few-body quantum phase
transitions \cite{Liu21, Ying22, Ying22new}, a topic which is arousing the interest of a growing number of researchers \cite{Hwang15, Hwang16, Ying15, Liu17, Ying21}.
The emergence of such peculiar transitions in the matter-radiation interaction is strongly related to the possibility of achieving the strong \cite{Wallraff04}, ultrastrong \cite{Forn16, Baust16, Forn19}, and deep strong coupling regimes \cite{Casanova10, Forn17, Yoshihara17}.

The quantum Rabi model (QRM) \cite{Rabi36, Rabi37, JC} is the simplest paradigmatic few-body system exhibiting the occurrence of critical phenomena \cite{Bakemeier12, Ashhab13, Liu21, Ying22, Ying22new, Hwang15, Hwang16, Ying15, Liu17, Ying21}.
In the last three decades the QRM has been implemented in different regions of its parameter space proving to be  an adaptable theoretical resource \cite{Hines04, Levine04, Ashab10, Hwang10} to investigate several radiation-matter scenarios as for example trapped ions \cite{Puebla16, Puebla17}, cavity QED \cite{Raimond01} and circuit QED systems \cite{Yang17, Baksic14, Xie14}.

The dynamical behaviours \cite{Hwang15, Hwang16, Shen17} as well as the phase diagrams \cite{Baksic14} exhibited by the QRM and the Dicke model \cite{Dicke54}, suggest the possible existence of a correspondence between the few-body and the regular QPTs.
By analysing the scaling of the critical exponents, the QPT exhibited by the QRM can be indeed connected to many-body and thermodynamic cases \cite{Liu17}.
When one deals with the QRM, the standard thermodynamic limit must be \textit{de facto} replaced by the so called classical oscillator limit \cite{Bakemeier12, Ashhab13}, which consists in the ideal physical regime identified by the vanishing spin-to-field frequency ratio.
It is worth noticing that the QRM presents critical phenomena also in the finite-frequency regime \cite{Liu21, Ying22, Ying22new} and recently a Beretzinski-Kosterlitz-Thouless QPT was found to occur in a dissipative QRM \cite{DeFilippis23}.
These properties of the QRM thus spur to search for new physical scenarios where few-body systems can manifest, in special parameter space regions, intriguing critical phenomena conceptually traceable back to the standard QPTs in the thermodynamic limit.

It seems therefore reasonable to foresee a comparable dynamic richness for the extended and generalized versions of the QRM, like the two-qubit \cite{Agarwal12, Peng12, Lee13, Chilingaryan13, Wang14, Peng14} and multi-qubit \cite{Peng21, Zhang21prl} QRM, the two-photon \cite{Chen12, Felicetti15} and multi-photon \cite{Zhang13} QRM, and the multi-level QRM \cite{Albert12, giannelli23prr}.
Recently, investigating a two-qubit QRM where a non-trivial qubit-qubit interaction is considered, it has been  brought to light a first-order QPT in the finite-frequency limit \cite{GdCMSV}.
%Until now the `standard' two-qubit QRMs analysed \cite{Zhang15, Duan15, Dong16, Mao19, Sun20, Yan21, Zhang21, Liu21, Mao21} have presented no qubit-qubit interaction.
We remark that introducing a two-qubit coupling in the Hamiltonian model complies with the main goal of quantum computation and quantum information, namely the implementation of two-qubit quantum logic gates \cite{Kang16, Lu13, Li09}.

In this letter we deal with an integrable and exactly solvable two-qubit QRM.
%The integrability of the model allows to reduce the study of the dynamical problem into two independent sub-problems.
%The existence of a constant of motion induces indeed the existence of two invariant subspaces, where the dynamics of the two-qubit QRM can be separately studied and solved.
It is possible to clearly demonstrate that, crossing the critical value of an appropriately defined adimensional control parameter, the tripartite (qubit-qubit-radiation mode) system undergoes a second-order superradiant QPT when an appropriately defined adimensional control parameter crosses a physically interpretable critical value.
%For small values of the control parameter the system is characterized by a phase where the ground state is undetermined since it is delocalized between the two invariant orthogonal subspaces.
%After crossing the critical point, the ground state localizes in one subspace, and a superradiant quantum phase transition, characterized also by a drastic change in the qubit-qubit entanglement behaviour, occurs.
The peculiarity of such a QPT lies in the nature of the classical limit involved.
In this case, indeed, what goes to infinity is the ratio of both the qubit-qubit and the qubit-mode couplings to the oscillator frequency.
The frequencies of the qubits remain instead free parameters and then can assume values close to the oscillator's frequency.
A new physical condition for reaching the thermodynamic-like limit is then brought to light in the framework of the adopted two-qubit QRM.

\textit{Model.}
Consider the following Hamiltonian model (in units of $\hbar$):
\begin{equation} \label{Hamiltonian}
\begin{aligned}
{H} = &
\omega \hat{a}^\dagger \hat{a}
+ \epsilon_1 \hat{\sigma}_1^z + \epsilon_2 \hat{\sigma}_2^z
+ \gamma\hat{\sigma}_{1}^{x}\hat{\sigma}_{2}^{x}
+ (\lambda_1 \hat{\sigma}_1^z + \lambda_2 \hat{\sigma}_2^z) (\hat{a} +\hat{a}^\dagger) ,
\end{aligned}
\end{equation}
which describes two interacting, biased ($\epsilon_1$ and $\epsilon_2$) qubits coupled to a single bosonic field mode.
$\omega$ is the characteristic frequency of the mode, while $\gamma$ and $\lambda_i$ ($i=1,2$) are the spin-spin and the ($i$-th) spin-mode couplings, respectively.
$\hat{\sigma}_{k}^{l}$ ($k=1,2$, $l=x,y,z$) are qubit Pauli operators, while $a$ and $a^\dagger$ are the annihilation and creation boson operators, respectively.

Since $\hat{\sigma}_{1}^{z}\hat{\sigma}_{2}^{z}$ is a constant of motion, the Hamiltonian can be unitarily transformed into $H = H_+ \oplus H_-$, with (see supplemental material)
\begin{equation} \label{H+ H-}
\begin{aligned}
{H}_{\pm} = &
\omega ~ \hat{a}^\dagger \hat{a} +
\epsilon_\pm \hat{\sigma}^z +
\gamma \hat{\sigma}^{x} +
\lambda_\pm \left( \hat{a}^\dagger + \hat{a} \right)  \hat{\sigma}^z,
\end{aligned}
\end{equation}
where $\epsilon_\pm = \epsilon_1 \pm \epsilon_2$, $\lambda_\pm = \lambda_1 \pm \lambda_2$ and $\hat{\sigma}^l$ ($l=x,y,z$) are standard two-level Pauli operators.
$H_+$ ($H_-$) is the effective Hamiltonian governing the dynamics of the two-qubit-mode system within the dynamically invariant subspace $\mathcal{H}_+$ ($\mathcal{H}_-$), which is spanned by $\{ \ket{\uparrow\uparrow},\ket{\downarrow\downarrow} \} \otimes \{ \ket{n} \}_{n \in 0}^\infty$ ($\{ \ket{\uparrow\downarrow},\ket{\downarrow\uparrow} \} \otimes \{ \ket{n} \}_{n \in 0}^\infty$), where we defined $\hat{\sigma}^z\ket{\uparrow}=+\ket{\uparrow}$, $\hat{\sigma}^z\ket{\downarrow}=-\ket{\downarrow}$, and $\hat{a}^\dagger \hat{a}\ket{n}=n\ket{n}$ \cite{GdCMSV, GVSM}.
We remark that $H_+$ and $H_-$ result to be single-qubit QRM Hamiltonians, where the qubit-qubit coupling $\gamma$ plays the role of the strength of a fictitious  transverse field.

It must be noted that two qubits holistically behave as a two-level system within each invariant subspace, meaning that $\hat{\sigma}^l$ ($l=x,y,z$) are the Pauli operators of such an effective (fictitious) two-level system.
Thus, the dynamics of the original system (two interacting qubits coupled to the same quantized field mode) can be exactly reduced to that of two independent effective single-spin quantum Rabi problems.
We can therefore treat the original dynamical problem by solving the single-spin QRMs defined in $\mathcal{H}_+$ and $\mathcal{H}_-$.
%In particular, thanks to such an exact dynamical decomposition, the eigenvalue problem can be successfully dealt with.
%The spectrum of the two-spin-mode system can be obtained by the `union' of the spectra of $H_+$ and $H_-$.
%Moreover, there exist  physically easily interpretable conditions on the parameters appearing in the reduced  Hamiltonian models making the eigenvalue problems exactly  solvable in a trivial way \cite{GdCMSV}.

It is worth noticing that, since such a reduction of $\mathcal{H}$ into $\mathcal{H}_+$ and $\mathcal{H}_-$ is independent of the Hamiltonian parameters, this approach keeps then its validity in the weak, strong, ultra-strong and deep-strong regimes (see Refs. \cite{Forn19, Kockum19, Xie17} for the classification of the spin-mode coupling regimes).
Moreover we underline that  such a reduction can be  exactly achieved even in the case of time-dependent controlling classical fields, that is $\epsilon_1(t)$ and $\epsilon_2(t)$ \cite{GGMIM, GMMM, GdCNM2}.

\textit{Biased QPT.}
The parameters $\epsilon_\pm$ and $\lambda_\pm$, characterizing the two effective Hamiltonians \eqref{H+ H-}, suggest to consider the case of counter-biased qubits, i.e. $\epsilon_1=-\epsilon_2=\epsilon/2$, equally coupled to the field mode, namely $\lambda_1=\lambda_2=\lambda/2$.
We stress that such a particular scenario relative to the Hamiltonian model \eqref{Hamiltonian}, simply corresponds to  realistic internal geometric symmetries governing the coupling in the tripartite system and demands an external classical control quite easy to experimentally implemented.
The advantage of such particular conditions is that the two effective Hamiltonians \eqref{H+ H-} assume the simple forms
\begin{equation} \label{H+ H- new}
\begin{aligned}
{H}_{+} = &
\omega ~ \hat{a}^\dagger \hat{a} +
\gamma \hat{\sigma}^{x} +
\lambda \left( \hat{a}^\dagger + \hat{a} \right)  \hat{\sigma}^z,
\\
{H}_{-} = &
\omega ~ \hat{a}^\dagger \hat{a} +
\epsilon \hat{\sigma}^z +
\gamma \hat{\sigma}^{x}.
\end{aligned}
\end{equation}
We see that $H_+$ possesses the form of the standard QRM Hamiltonian, while $H_-$ describes a (fictitious) two-level system decoupled from a bosonic field mode.

Let us now analyze the particular regime reached by the tripartite system in the region of the parameter space asymptotically defined by the following two conditions:
\begin{equation} \label{limits}
\begin{aligned}
\gamma/\omega \rightarrow \infty,
\qquad
\lambda/\omega \rightarrow \infty.
\end{aligned}
\end{equation}
These limits ensure the classical oscillator limit \cite{Bakemeier12, Hwang15}, which resembles the role that the thermodynamic limit plays in many-body spin systems \cite{Bakemeier12, Ashhab13}.
Such a kind of thermodynamic-like limit has been introduced for the single-qubit QRM \cite{Bakemeier12}.
In that case, however, it is the ratio of the spin frequency to that of the oscillator which individuates the classical limit regime.
In our case, instead, the crucial ratios are the spin-spin and the spin-mode couplings to the oscillator frequency ($\gamma/\omega$ and $\lambda/\omega$, respectively), regardless of the magnitude of the qubit's frequency.

It is possible to verify that, under the conditions individuated in Eq. \eqref{limits}, the exact low-energy form of the Hamiltonian $H_+$ is (see the supplemental material)
\begin{equation} \label{Ham np}
\begin{aligned}
{H}_{+}^{np} =
\omega ~ \hat{a}^\dagger \hat{a} -
{\omega g^2 \over 4} (\hat{a}^\dagger + \hat{a})^2 - \gamma,
\end{aligned}
\end{equation}
for $g<1$, and
\begin{equation} \label{Ham sp}
\begin{aligned}
{H}_{+}^{sp} =
\omega ~ \hat{a}^\dagger \hat{a} -
{\omega \over 4 g^4} (\hat{a}^\dagger + \hat{a})^2 -
\gamma ~ {g^2 + g^{-2} \over 2},
\end{aligned}
\end{equation}
for $g>1$, with $g=\sqrt{2}\lambda/\sqrt{\omega\gamma}$.
The fictitious single-qubit QRM $H_+$ in Eq. \eqref{H+ H- new} exhibits a second-order QPT \cite{Hwang15}, guided by the parameter $g$, between a normal ($np$) and a superradiant ($sp$) phase.
Here we can appreciate the merit of the exact  subdivision of the Hilbert space into the two dynamically invariant subspaces $\mathcal{H}_+$ and $\mathcal{H}_-$.
Thanks to such a dynamical decomposition and the reinterpretation of the two subdynamics in terms of fictitious single qubit-mode systems, we can in fact transparently formulate the occurrence of a QPT also in our tripartite two-qubit QRM too.
It is therefore worth seeking the effects on the ground state and its physical properties traceable back to the normal-superradiant QPT.

The supplemental material briefly sketches how to calculate the lowest-energy states and the corresponding eigenenergies of $H_+^{np}$ and $H_+^{sp}$.
Basing on the two-qubit - single-qubit mapping, such states, in terms of the two-qubit-mode system, read:
\begin{equation} \label{GS np}
\begin{aligned}
E_{0+}^{np} &= \omega {\sqrt{1-g^2} - 1 \over 2} - \gamma,
\\
\ket{\Psi_{0+}^{np}} &= \mathcal{S}[r_{np}(g)] \ket{0} \otimes {\ket{\uparrow\uparrow}-\ket{\downarrow\downarrow} \over \sqrt{2}},
\end{aligned}
\end{equation}
for $g<1$, and
\begin{equation} \label{GS sp}
\begin{aligned}
E_{0+}^{sp} =& ~ \omega {\sqrt{1-g^2} - 1 \over 2} - \gamma{g^2 + g^{-2} \over 2},
\\
\ket{\Psi_{0+}^{sp}} =& ~ \mathcal{S}[r_{sp}(g)] \ket{0} \otimes \\
& \Bigg[ \left( {\sqrt{1+g^{-2}} \mp \sqrt{1-g^{-2}} \over 2} \right) \ket{\uparrow\uparrow} - \\
& \hspace{1cm} \left( {\sqrt{1+g^{-2}} \pm \sqrt{1-g^{-2}} \over 2} \right) \ket{\downarrow\downarrow} \Bigg],
\end{aligned}
\end{equation}
when $g>1$.
Here $r_{np} = - \ln(1-g^2)/4$, $r_{sp} = - \ln(1-g^{-4})/4$, and $\mathcal{S}(x) = \exp\{ (x/2) (\hat{a}^{\dagger2} - \hat{a}^2) \}$.
The double sign means that the ground state of $H^{sp}_+$ is two-fold degenerate.

\begin{figure}[] 
\begin{center}
\includegraphics[width=0.45\textwidth]{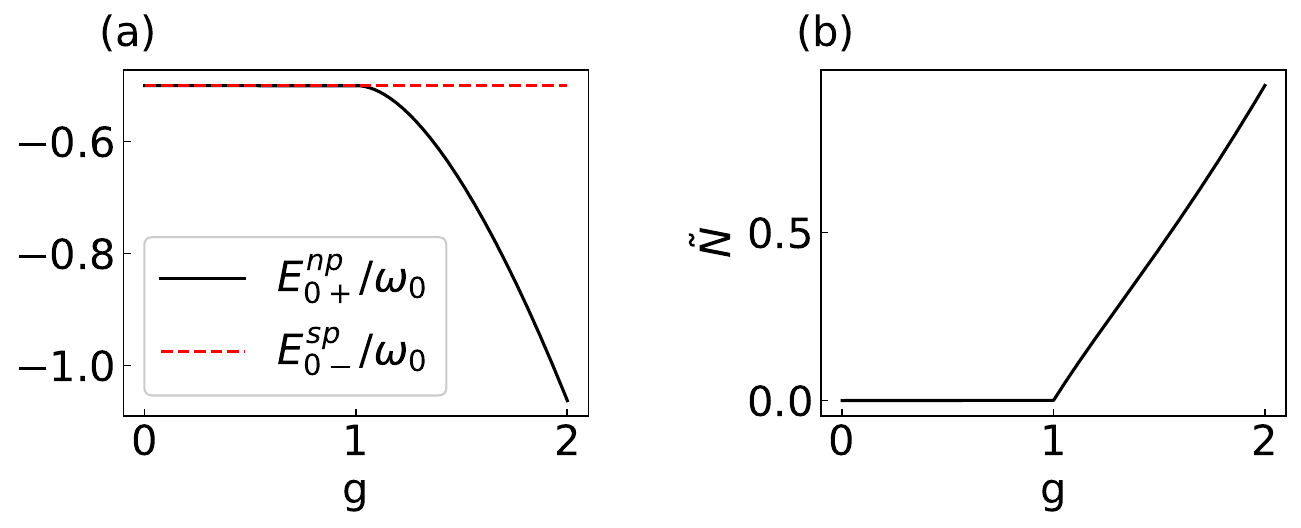}
\captionsetup{justification=raggedright,format=plain,skip=4pt}%
\caption{Dependence of the first two lowest eigenenergies $\widetilde{E}_0^+/\omega$ and $\widetilde{E}_0^-/\omega$ on the control parameter $g=g_1=g_2$, for counter-biased qubits ($\epsilon_1=-\epsilon_2)$ and equal spin-mode couplings ($\lambda_1=\lambda_2$).
The black curve in the superradiant phase ($g>1$) corresponds to two degenerate eigenstates (see Eq. \eqref{GS sp}).
(b) Dependence of the mean photon number on $g$, for the ground state of the two-qubit QRM, with $\epsilon / \gamma \rightarrow 0$, $\omega / \gamma \rightarrow 0$, and $\omega / \lambda \rightarrow 0$.}
\label{fig: qrmII_multiplot_1}
\end{center}
\end{figure}

Let us suppose that the qubits' energy is comparable with that of the oscillator, i.e. $\epsilon \ll \gamma$.
In such a condition the lowest energy of $H_-$ and the related eigenstate can be easily derived; the corresponding eigenstate and the related energy of the tripartite system read
\begin{equation} 
\begin{aligned}
E_{0}^{-} &= - \gamma % \sqrt{\gamma^2 + \epsilon^2},
\\
\ket{\Psi_{0}^{-}} &= \ket{0} \otimes {\ket{\uparrow\downarrow}-\ket{\downarrow\uparrow} \over \sqrt{2}}.
\end{aligned}
\end{equation}
If we introduce the rescaled energy $\widetilde{E}=E \cdot \omega/\gamma$, in the limit $\gamma/\omega \rightarrow \infty$we get
\begin{equation} \label{rescaled energies}
\begin{aligned}
\widetilde{E}_0^+ &=
\begin{cases}
\widetilde{E}_{0+}^{np} = - \omega, \quad g<1
\\\\
\widetilde{E}_{0+}^{sp} = - \omega ~ ({g^2 + g^{-2}) / 2}, \quad g>1
\end{cases}
\\\\
\widetilde{E}_{0}^{-} &= - \omega.
\end{aligned}
\end{equation}
We see that for $g<1$ the ground state of the tripartite system has a double degeneracy and, in general, it possesses nonvanishing projections in $\mathcal{H}_+$ and $\mathcal{H}_-$.
%{\color{red}In this instance the ground state is then `delocalized' between the two invariant subspaces $\mathcal{H}_+$ and $\mathcal{H}_-$.}
For $g>1$, instead, the ground state of the same system is $\ket{\Psi_{0+}^{sp}}$ and belongs to the subspace $\mathcal{H}_+$ since $\widetilde{E}_{0+}^{sp} < \widetilde{E}_0^-$, as clearly shown in Fig. \ref{fig: qrmII_multiplot_1}(a), where the three rescaled energies in Eq. \eqref{rescaled energies} are plotted in units of $\omega$.

\begin{figure}[] 
\begin{center}
\includegraphics[width=0.45\textwidth]{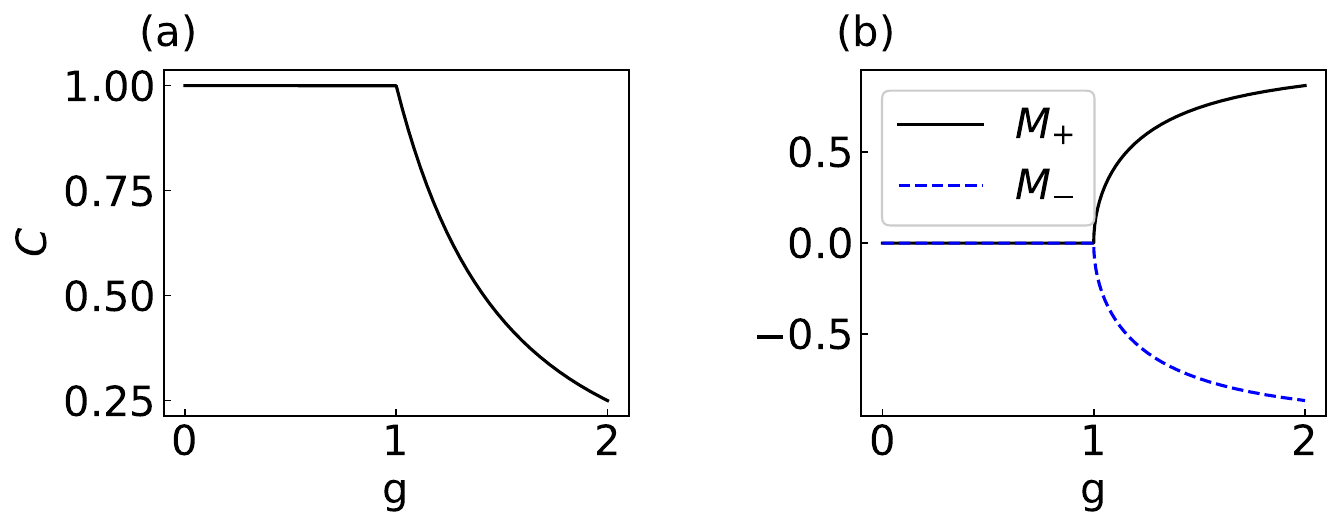}
\captionsetup{justification=raggedright,format=plain,skip=4pt}%
\caption{
Dependence of (a) the two-qubit concurrence, and (b) the two-qubit magnetization on $g$, for the ground state of the two-qubit QRM, with $\epsilon / \gamma \rightarrow 0$, $\omega / \gamma \rightarrow 0$, and $\omega / \lambda \rightarrow 0$.
The two curves (black and blue) for $g>1$ (superradiant phase) are related to the two states in Eq. \eqref{GS sp}
}
\label{fig: qrmII_multiplot_2}
\end{center}
\end{figure}

It is worth pointing out that, given the factorization of the ground states (for $g<1$ and for $g>1$, between qubit and bosonic subsystems), the mean photon number $N=\average{\Psi_0|\hat{a}^\dagger \hat{a}|\Psi_0}$, the two-qubit magnetization $M=\average{\Psi_0|(\hat{\sigma}_1^z+\hat{\sigma}_2^z)/2|\Psi_0}$ and the concurrence $C$ \cite{Wootters98} can be easily obtained ($\ket{\Psi_0}$ indicates here the generic ground state, independently of the phase, either normal or superradiant).
In the following these three quantities for both the normal ($g<1$) and the superradiant ($g>1$) phase are given,
\begin{equation} \label{physical quantities}
\begin{aligned}
\widetilde{N}_{np} = 0, &\qquad \widetilde{N}_{sp} = {g^2 - g^{-2} \over 4},\\
C_{np} = 1, &\qquad C_{sp} = g^{-2}, \\
M_{np} = 0, &\qquad M_{sp} = \pm \sqrt{1 - g^{-2}},
\end{aligned}
\end{equation}
where $\widetilde{N} = N \cdot (\omega/\gamma)$ is the rescaled mean photon number.
We underline that, in the case of the mean photon number, a rescaling is necessary to highlight the relative weight of the physical variable in the two phases, contrarily to what happens for the concurrence and the magnetization.
For $g>1$ the (not rescaled) mean photon number is indeed infinite as usual for the superradiant phase. 
The dependence of these three quantities on the dimensionless parameter $g$ is shown in Figs. \ref{fig: qrmII_multiplot_1}(b), \ref{fig: qrmII_multiplot_2}(a), and  \ref{fig: qrmII_multiplot_2}(b). 
We note that the mean two-qubit magnetization $M$ depends on the nature of the ground state in the superradiant phase.
The two curves (black and blue) in Fig. \ref{fig: qrmII_multiplot_2}(b) are indeed due to the degeneracy of the ground state in the superradiant phase [see Eq. \eqref{GS sp}].
This means that the associated statistical ensemble should be described by a density matrix with equal weights for the two states in Eq. \eqref{GS sp}, and then it would be characterized by a vanishing mean value of the two-qubit magnetization.

Conversely, the mean photon number and the concurrence have uniquely determined values in both the phases.
The drastic change of the dependence of the mean photon number on $g$ is at the origin of the superradiant nature of the QPT.
Moreover, the two phases are characterized by a different level of entanglement exhibited by the two qubits.
Precisely, the normal phase is characterized by a maximum level of entanglement ($C_{np}=1$), while the superradiant phase exhibits a concurrence which decreases as $g$ increases.
Therefore, besides the mean photon number, also the level of entanglement between the two qubits can serve as a signature of the occurrence of the superradiant QPT, like in other spin systems \cite{Osterloh02, Oliveira06, Gu07, Chen16, Yuste18}.
We thus see the occurrence of a second-order QPT from a normal to a superradiant phase for the two-qubit quantum Rabi system, which is characterized by two physical quantities ($N$ and $C$) which exhibit a critical behaviour: they are not derivable at the critical point of the control parameter ($g_c$) where the transition occurs.

\textit{Unbiased QPT}.
If we relax the constraints imposed to the parameters appearing in the Hamiltonian model \eqref{Hamiltonian}, and consider the more general case $|\epsilon_1| \neq |\epsilon_2|$, a bias term (namely $\epsilon_+ \hat{\sigma}^z$) would appear in $H_+$.
We emphasize that, in this instance, this more general Hamiltonian cannot be approximated by the expressions in Eqs. \eqref{Ham np} and \eqref{Ham sp} in the two regimes.

\begin{figure}[] 
\begin{center}
{\includegraphics[width=0.35\textwidth]{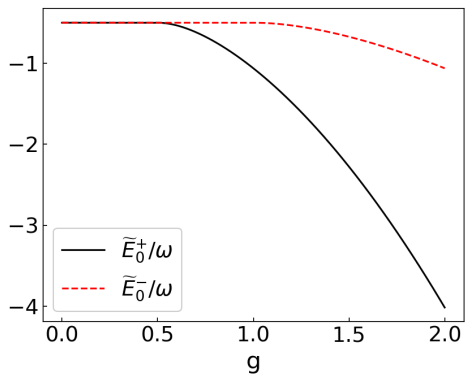}}
\captionsetup{justification=raggedright,format=plain,skip=4pt}%
\caption{Dependence of the first two lowest eigenenergies $\widetilde{E}_0^+/\omega$ and $\widetilde{E}_0^-/\omega$ on the control parameter $g=g_-=g_+/2$ (i.e, $g=2g_1/3=2g_2$), for unbiased qubits ($\epsilon_1=\epsilon_2=0)$ and different spin-mode couplings ($\lambda_1=3\lambda_2$, implying $\lambda_+/2=\lambda_-=\lambda$).}
\label{fig: qrmII_energies}
\end{center}
\end{figure}

In light of the previous observation, we change the physical scenario assuming no magnetic field(s) acting upon the two qubits, namely $\epsilon_1=\epsilon_2=0$ (implying $\epsilon_+=\epsilon_-=0$).
Such a  condition, even keeping $\lambda_1 \neq \lambda_2$, is still compatible with the approximation procedure leading to the two approximated Hamiltonians $H_+^{np}$ and $H_+^{sp}$ [Eqs. \eqref{Ham np} and \eqref{Ham sp}].
This time, the same approximation procedure can (and must) be analogously performed for $H_-$.
In this case the Hamiltonians $H_+^{np}$ ($H_+^{sp}$) and $H_-^{np}$ ($H_-^{sp}$) have a similar structure and it is possible to persuade oneself  that the same occurs for both the corresponding eigensolutions [see Eqs. \eqref{GS np} and \eqref{GS sp}].
The two lowest energies are plotted in Fig. \eqref{fig: qrmII_energies} for $\lambda_+/2 = \lambda_- = \lambda$, which implies $g_+/2 = g_- = g$ (with $g_\pm = \sqrt{2} \lambda_\pm / \sqrt{\omega\gamma}$).
We see that the normal-superradiant phase transition occurs in both the subspaces.
Nevertheless, the condition $\lambda_+ \neq \lambda_-$, which implies $g_+ \neq g_-$, produces a different dependence of the lowest energy level of $H_+^{sp}$ and $H_-^{sp}$ on the control parameter.
Therefore, in general, $g_c^+ \neq g_c^-$, with $g_c^\pm$ being the critical values of $g$ for which the QPT occurs in the subspace $\mathcal{H}_+$ and $\mathcal{H}_-$, respectively.
In the specific considered example, i.e. $\lambda_+/2 = \lambda_- = \lambda$ (implying $g_+/2 = g_- = g$), we have: $g_c^+=0.5$ and $g_c^-=1$ as shown in Fig. \eqref{fig: qrmII_energies}.
This fact implies that, in this instance, for $g<g_c^+$ the ground state of the two-qubit-mode system has nonvanishing projections on both $\mathcal{H}_+$ and $\mathcal{H}_-$, while for $g>g_c^+$ it uniquely belongs to the subspace $\mathcal{H}_+$.
Such a transition is still characterized by the physical quantities $C$, $M$ and $N$ which present the same dependence on $g$ seen before.
Finally, more in general, we can say that the subspace where the ground state is placed after the QPT depends on the quantity min$(g_c^+, g_c^-)$.

\textit{Remarks.}
%The occurrence of a second-order superradiant QPT in a two-qubit QRM is the first result reached in this letter.
%It is related to the existence of an analogous QPT in a single-qubit QRM scenario. 
%The link stems from the symmetries characterizing the two-qubit QRM, which  highlights that the dynamics of this tripartite system nests two independent fictitious single-qubit QRM scenarios. 
The central result of this work consists in the physical conditions given in Eq. \eqref{limits} that determine the thermodynamic limit under which a two-qubit QRM system undergoes a second-order superradiant QPT.
The standard thermodynamic limit in many-body systems \cite{Bakemeier12} is usually implemented by letting the number of spins go to infinity \cite{Dicke54}.
%Only in this limit the gap between the ground and the first excited state gets closed and the (second-order) QPT occurs \cite{Hepp73, Wang73, Emary03, Emary03E}.
However, it has been demonstrated that such a condition can be achieved also in finite-size systems \cite{Bakemeier12, Ashhab13}.
In particular, for the QRM the thermodynamic-like limit, in presence of a second-order QPT, results from pushing to infinity the spin-to-mode frequency ratio \cite{Bakemeier12, Hwang15}.

In this article, we have shown that in the two-qubit version of the QRM, such a thermodynamic limit emerges in a different way.
%We have indeed found that it is the ratio of the spin-spin coupling strength to the mode frequency what drives the two-qubit QRM system towards such a limit.
%This result can be physically interpreted by considering that the spin-spin coupling plays the role of the transverse field in the effective QRM Hamiltonians, which govern the two-qubit-mode dynamics within the two invariant subspaces.
We have indeed found that, in order to reach the classical oscillator limit \cite{Bakemeier12}, the ultrastrong interaction regime must be fulfilled by both the spin-spin and the spin-mode couplings ($\gamma/\omega \rightarrow \infty$, $\lambda/\omega \rightarrow \infty$), regardless of the spin-to-mode frequency ratio.
This circumstance highlights that in more complex systems with interacting qubits coupled to quantized field mode(s) there exist different opportunities (related to different parameters characterizing the model) to reach the thermodynamic limit.

For experimental implementation of the unveiled physics the best candidates are trapped-ion \cite{Dingshun18} and superconducting-circuit architectures \cite{Langford17}. Two trapped ions or superconducting qubit have been considered since long in a variety of academic and industrial labs \cite{Mills22, Moskalenko22, Zhao20}. The required vanishing of the field mode frequency in the QRM is a natural consequence of working in specific interaction pictures, as happened in the case of the simulation of the Dirac equation or the single-qubit QRM \cite{Gerritsma10}.

\textit{
This work was supported by the PNRR MUR project PE0000023-NQSTI.
GF and EP acknowledge the QuantERA grant SiUCs (Grant No. 731473) and the grant Pia.Ce.Ri.-UNICT, project Q-ICT.
}

\bibliography{biblio_qrm.bib}

\end{document}